\documentclass[aps,prd,preprintnumbers,amsmath,amssymb,nofootinbib,11pt]{revtex4}
\usepackage{eepic}
\usepackage{indentfirst}
\usepackage{mathrsfs}
\usepackage{fancyhdr}
\usepackage{ulem}
\usepackage{float}
\usepackage{graphicx}
\usepackage[
colorlinks=true, linkcolor=black, breaklinks=true, urlcolor=blue,
citecolor=green]{hyperref}
\usepackage{epstopdf}
\usepackage{bm,bbm}

\usepackage{xcolor}

\usepackage{tabularx}
\usepackage{subfigure}
\usepackage{epsfig}
\usepackage{color}
\usepackage{slashed}
\usepackage{hyperref}

\newcommand{\be}{\begin{equation}}
\newcommand{\ee}{\end{equation}}
\newcommand{\bea}{\begin{eqnarray}}
\newcommand{\eea}{\end{eqnarray}}
\renewcommand{\L}{\mathscr{L}}

\newcommand{\bra}{\langle}
\newcommand{\ket}{\rangle}
\newcommand{\nn}{\nonumber}
\newcommand{\MeV}{\,\text{MeV}}
\newcommand{\GeV}{\,\text{GeV}}
\renewcommand{\vec}[1]{\mathbf{#1}}
\newcommand{\diff }{{\text{d}}}

\newcommand{\disc}{\text{disc}\,}
\newcommand{\nl}{\notag\\}

\newcommand{\beq}{\begin{equation}}
\newcommand{\eeq}{\end{equation}}
\newcommand{\bsp}{\begin{sloppypar}}
\newcommand{\esp}{\end{sloppypar}}

\begin{document}
\pagestyle{plain}

\title {\boldmath Unified study of $B_s^0 \to X(3872) \pi^+\pi^- (K^+ K^-)$ and $B_s^0 \to \psi(2S) \pi^+\pi^- (K^+ K^-)$ processes}
\author{ Yun-Hua~Chen}\email{yhchen@ustb.edu.cn}
\affiliation{School of Mathematics and Physics, University
of Science and Technology Beijing, Beijing 100083, China}

\begin{abstract}

We perform a unified description of the experimental data of the $\pi^+\pi^-$ invariant mass spectra of $B_s^0 \to \psi(2S) \pi^+\pi^-$, the $\pi^+\pi^-$ and $K^+ K^-$ invariant mass spectra of $B_s^0 \to X(3872) \pi^+\pi^- (K^+ K^-)$, and the ratio of branch fractions $\mathcal{B}[B_s^0 \to X(3872)(K^+K^-)_{{\rm non-}\phi}]/ \mathcal{B}[B_s^0 \to X(3872)\pi^+ \pi^-)$. The strong final state interactions between the two pseudoscalars are taken into account using a parametrization fulfilling unitarity and analyticity. We find that there is universality in the coupling constants for $B_s^0 \to \psi(2S) \pi^+\pi^-$ and
$B_s^0 \to J/\psi \pi^+\pi^-$ processes. While the couplings of $B_s^0 \to X(3872) \pi^+\pi^-$ are about half of magnitude smaller than the couplings of $B_s^0 \to \psi(2S) \pi^+\pi^-$, which indicates that the $X(3872)$ is different from a pure charmonium state. Furthermore, we find that the $f_0(1500)$ plays an important role in the $B_s^0 \to \psi(2S) \pi^+\pi^-$ and the $B_s^0 \to X(3872) \pi^+\pi^- (K^+ K^-)$ processes, though the phase space of $B_s^0 \to X(3872) f_0(1500)$ is small. Also we predict the ratio of branch fractions $\mathcal{B}[B_s^0 \to \psi(2S)(K^+K^-)_{{\rm non-}\phi}]/ \mathcal{B}[B_s^0 \to \psi(2S)\pi^+ \pi^-]$ and the $K^+ K^-$ invariant mass distribution of $B_s^0 \to \psi(2S)K^+K^-$.

\end{abstract}

\maketitle

\newpage
\section{Introduction}

The nature of the vector charmoniumlike state $X(3872)$, also known as $\chi_{c1}(3872)$, has remained controversial
since its discovery in the $J/\psi\pi^+\pi^-$ invariant mass spectrum from
$B$ meson decays in 2003 by the Belle Collaboration~~\cite{Belle:2003nnu}. It was subsequently
confirmed by the analyses of the data of the $p\bar{p}$ collisions by
the CDF~\cite{CDF:2003cab} and D$0$~\cite{D0:2004zmu} Collaborations. The mass of the X(3872)
almost exactly coincides with the threshold of $D^0 \bar{D}^{\ast 0}$, and it has large decay rate to $D^0 \bar{D}^{\ast 0}$. Another important feature of the $X(3872)$ is that the large isospin violating effects observed in its decay patterns. Models have been proposed to interpret the $X(3872)$ as a $D^\ast \bar{D}/\bar{D}^\ast D$ molecule~\cite{boundstate1,boundstate2,boundstate3,boundstate4,Wang:2017dcq,virtualstate,Baru:2024ptl}, a compact tetraquark state~\cite{tetraquark}, a $\chi_{c1}(2P)$ state~\cite{Zhang:2009bv,Meng:2014ota}, a hybrid state~\cite{hybrid}, the mixing of the $c\bar{c}$ core with the $D^\ast \bar{D}/\bar{D}^\ast D$~\cite{ccbar1,ccbar2} and so on. See Refs.~\cite{Chen:2016qju,Hosaka:2016pey,Lebed:2016hpi,Esposito:2016noz,Guo:2017jvc,Ali:2017jda,Olsen:2017bmm,Karliner:2017qhf,Kalashnikova:2018vkv,Brambilla:2019esw,Meng:2022ozq,Liu:2024uxn,Chen:2024eaq} for recent reviews.

Decays of beauty hadrons to final states with the $X(3872)$ provide a unique laboratory to
study its property, especially a comparison of its production rates with respect to those final states with conventional charmonium states can help to reveal the $X(3872)$'s internal structure~\cite{Maiani:2017kyi}. Recently, the $B^0_s \to X(3872) \pi^+ \pi^-$ decay was firstly observed by the LHCb collaboration~\cite{LHCb:2023reb}, in which the $B^0_s \to \psi(2S) \pi^+ \pi^-$ process was also measured. It is found that the dipion mass spectra in $B^0_s \to X(3872) \pi^+ \pi^-$ decay shows a similarity with those in $B^0_s \to \psi(2S) \pi^+ \pi^-$ decay. In this work, we will study the $B^0_s \to X(3872) \pi^+ \pi^-$ and $B^0_s \to \psi(2S) \pi^+ \pi^-$ processes in an unified scheme and extract relevant couplings. Given the almost pure $s\bar{s}$ source the pions are generated from,
the data of $B_s^0 \to X(3872) K^+ K^-$ process~\cite{LHCb:2020coc} and the ratio of branch fractions $\mathcal{B}[B_s^0 \to X(3872)(K^+K^-)_{{\rm non-}\phi}]/ \mathcal{B}[B_s^0 \to X(3872)\pi^+ \pi^-]$~\cite{ParticleDataGroup:2024cfk} will also be taken into account.

In $B_s^0 \to X(3872) [\psi(2S)] \pi^+\pi^- (K^+ K^-)$ processes, the two pseudoscalars invariant mass reaches up to about 1.5 GeV, so the coupled-channel final-state
interaction (FSI) in the $S$ wave is strong and needs to be taken into account properly. We will use the method developed in Ref.~\cite{Ropertz:2018stk} to consider the strong FSI, which marries the advantages
of a dispersive description at low energies below 1 GeV with the phenomenological model consistent with analyticity beyond. In Ref.~\cite{Ropertz:2018stk}, this method has been successfully employed to study of $B_s^0 \to J/\psi \pi^+\pi^- $ and $B_s^0 \to J/\psi K^+ K^-$ process. Here we perform a simultaneous analysis of the $B_s^0 \to X(3872) [\psi(2S)] \pi^+\pi^- (K^+ K^-)$ processes. At low energies, the amplitude should agree with the leading chiral results, and therefore we construct the chiral contact Lagrangian of $B_s^0  X(3872) [\psi(2S)] \pi^+\pi^- (K^+ K^-)$ couplings in the spirit of the chiral and the heavy-quark nonrelativistic expansions~\cite{Mannel,Chen2016,Chen:2019gty,Chen:2019mgp,Chen:2021aud}.

This paper is organized as follows. In Sec.~\ref{theor}, we present
the theoretical framework and elaborate on the calculation of the
amplitudes as well as the treatment of the FSI. In
Sec.~\ref{pheno}, we show our fit results and discussions, followed by a summary in Sec.~\ref{conclu}.

\section{Theoretical framework}\label{theor}
\subsection{ Lagrangians of contact $ B_s^0 \psi(2S) \pi^+\pi^- (K^+ K^-) $ couplings}

\begin{figure}
\centering
\includegraphics[height=6cm,width=10cm]{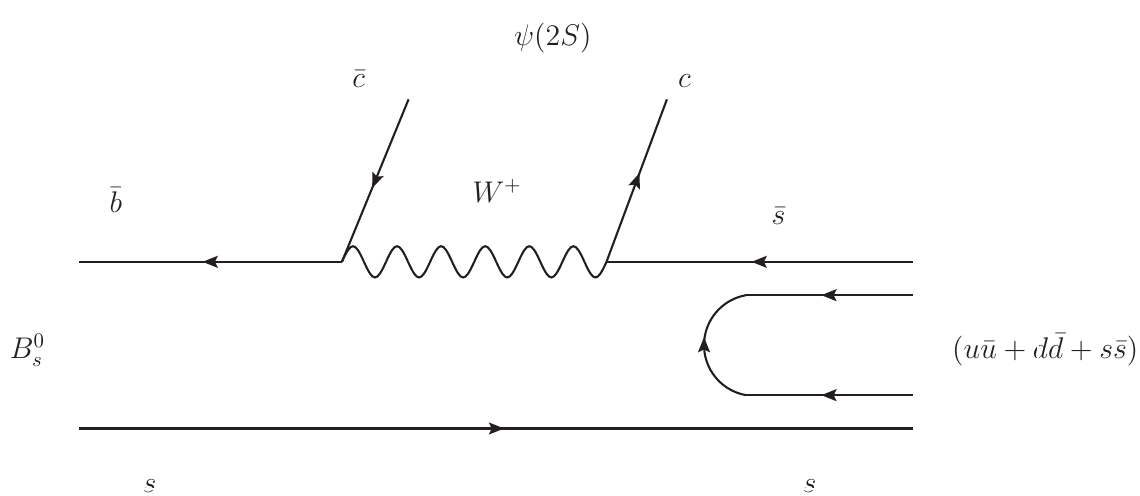}
\caption{The $ B_s^0 \to \psi(2S) \pi^+\pi^- (K^+ K^-) $ diagram to leading order via $W^+$ exchange.
}\label{fig.WeakDecay}
\end{figure}

The leading quark-level Feynman diagram contributing to $ B_s^0 \to \psi(2S) \pi^+\pi^- (K^+ K^-) $ is depicted in Fig.~\ref{fig.WeakDecay}, which can be separated into two steps.
First, $\bar{b}$ quark decays into a $\bar{c}$ quark and a $W^+$ boson followed by its decay into a $c$ quark and a $\bar{s}$ quark. Then, the $\psi(2S)$ is produced from the hadronization of $c\bar{c}$, and the pions (kaons) are generated from the $s\bar{s}$ source. If only the light-quark part is concerned, $B_s^0$ ``effectively'' provides the $s\bar{s}$ source in these decays.
The $s\bar{s}$ can be decomposed into SU(3) singlet and octet components,
\be
\label{eq.sbarsComponents}  s\bar{s}=\frac{\sqrt{3}}{3}|P_1\rangle-\frac{\sqrt{6}}{3}|P_8\rangle\,,
\ee
where $|P_1\rangle \equiv
\frac{1}{\sqrt{3}}(u\bar{u}+d\bar{d}+s\bar{s})$ and $|P_8\rangle \equiv
\frac{1}{\sqrt{6}}(u\bar{u}+d\bar{d}-2
s\bar{s})$.
Therefore, the effective Lagrangian for the $B_s^0\psi(2S)\pi\pi$ and
$B_s^0\psi(2S) K\bar{K}$ contact couplings, at leading order in the chiral as well as the heavy-quark nonrelativistic expansion,
reads~\cite{Mannel,Chen2016,Chen:2019gty,Chen:2019mgp,Chen:2021aud}
\begin{equation}\label{LagrangianBspsi2Spipi}
\L_{B_s^0\psi^\prime\Phi\Phi} = g_1\bra P_{1} \partial_i J^{i\dag} \ket \bra u_\mu
u^\mu\ket - \sqrt{2}g_1\bra \partial_i J^{i\dag} \ket \bra P_8 u_\mu
u^\mu\ket  +h_1\bra P_{1} \partial_i J^{i\dag} \ket \bra u_\mu
u_\nu\ket  v^\mu v^\nu-\sqrt{2}h_1\bra \partial_i J^{i\dag} \ket \bra  P_{8} u_\mu
u_\nu\ket  v^\mu v^\nu
+\mathrm{H.c.}\,,
\end{equation}
where $\langle\ldots\rangle$ denotes the trace in the SU(3) flavor space, $J= (\psi^\prime/\sqrt{3}) \cdot \mathbbm{1}$, and
$v^\mu=(1,\vec{0})$ is the velocity of the initial state heavy quark. In Eq.~\eqref{LagrangianBspsi2Spipi}, the Lagrangian are constructed by placing the SU(3) singlet parts and the SU(3) octet parts into different SU(3) flavor traces. Note that in $B_s^0\psi(2S)\pi\pi(K \bar{K})$ decays, a $p$-wave interaction is needed to match the angular momentum conservation.

The SU(3) octet of the Goldstone bosons of the spontaneous breaking of chiral symmetry can
be parametrized as
\begin{align}
u_\mu &= i \left( u^\dagger\partial_\mu u\, -\, u \partial_\mu
u^\dagger\right) \,, \qquad
u = \exp \Big( \frac{i\Phi}{\sqrt{2}F} \Big)\,,\nn\\
\Phi &=
 \begin{pmatrix}
   {\frac{1}{\sqrt{2}}\pi ^0 +\frac{1}{\sqrt{6}}\eta _8 } & {\pi^+ } & {K^+ }  \\
   {\pi^- } & {-\frac{1}{\sqrt{2}}\pi ^0 +\frac{1}{\sqrt{6}}\eta _8} & {K^0 }  \\
   { K^-} & {\bar{K}^0 } & {-\frac{2}{\sqrt{6}}\eta_8 }  \\
 \end{pmatrix} . \label{eq:u-phi-def}
\end{align}
Here $F$ is the pion decay constant in the chiral limit, and we use the physical value $92.1\MeV$ for it.

\subsection{ Lagrangians of contact $ B_s^0 X(3872) \pi^+\pi^- (K^+ K^-) $ couplings}

In contrast with the charmonium state $\psi(2S)$ which is a light-quark SU(3) flavor-singlet, the structure of the $X(3872)$ has remained controversial. Considering the possible light quarks it may contain (e.g. in the charm and anticharm mesons molecule or the fourquark scenarios), the $X(3872)$ can be decomposed into SU(3) singlet and octet components of light quarks,
\be
\label{eq.YComponents} |X(3872)\rangle=a|V_1\rangle+b|V_8\rangle\,,
\ee
where $|V_1\rangle \equiv
\frac{1}{\sqrt{3}}(u\bar{u}+d\bar{d}+s\bar{s})\otimes
V^{\text{heavy}}$ and $|V_8\rangle \equiv
\frac{1}{\sqrt{6}}(u\bar{u}+d\bar{d}-2
s\bar{s})\otimes V^{\text{heavy}}$.
Note that the heavy quark (e.g., $c$ quark) components are contained in the $V^{\text{heavy}}$, and they are not
distinguishable in $|V_1\rangle$ and $|V_8\rangle$. If the $X(3872)$ contains no light-quark (e.g. in the pure $c \bar{c}$ or the hybrid state scenarios), it is a light-quark SU(3) flavor-singlet. 
The effective Lagrangian for the $X(3872)\psi(2S)\pi\pi$ and
$X(3872)\psi(2S) K\bar{K}$ contact couplings, at lowest order in the chiral as well as the heavy-quark nonrelativistic expansion,
reads~\cite{Mannel,Chen2016,Chen:2019gty,Chen:2019mgp,Chen:2021aud}
\begin{eqnarray}\label{LagrangianBsXpipi}
\L_{B_s^0 X(3872)\Phi\Phi} = && \tilde{g}_1\bra P_{1} \partial_i X^{i\dag}_1 \ket \bra u_\mu
u^\mu\ket - \sqrt{2}\tilde{g}_1\bra \partial_i X^{i\dag}_1 \ket \bra P_8 u_\mu
u^\mu\ket  +\tilde{g}_8\bra P_{1}  \ket \bra \partial_i X^{i\dag}_8 u_\mu
u^\mu\ket-3\sqrt{2}\tilde{g}_8\bra P_{8}   \partial_i X^{i\dag}_8 u_\mu
u^\mu\ket   \nn \\
&& + \tilde{h}_1\bra P_{1} \partial_i X^{i\dag}_1 \ket \bra u_\mu
u_\nu\ket  v^\mu v^\nu - \sqrt{2}\tilde{h}_1\bra \partial_i X^{i\dag}_1 \ket \bra P_8 u_\mu
u_\nu\ket  v^\mu v^\nu +\tilde{h}_8\bra P_{1}  \ket \bra \partial_i X^{i\dag}_8 u_\mu
u_\nu\ket  v^\mu v^\nu  \nn \\
&& -3\sqrt{2}\tilde{h}_8\bra P_{8}   \partial_i X^{i\dag}_8 u_\mu
u_\nu\ket  v^\mu v^\nu
+\mathrm{H.c.}\,.
\end{eqnarray}
If we only consider the short-range contribution to the hadronization of $c\bar{c}$, the mechanism for the production of the $ X(3872)$ is the same as that shown in Fig.~\ref{fig.WeakDecay}. The difference of the long-distance contribution to the productions of the $\psi(2S)$ and the $ X(3872)$ is reflected in the different Lagrangian structures in Eqs.~\eqref{LagrangianBspsi2Spipi} and~\eqref{LagrangianBsXpipi}, and in the different values of the coupling constants.

\subsection{Amplitudes of \boldmath{$ B_s^0 \to \psi(2S)[X(3872)] \pi^+\pi^- (K^+ K^-) $} processes} \label{subsectionC}

First we define the Mandelstam variables in the decay
process of $ B_s^0(p_a) \to
Y(p_b) P(p_c)P(p_d) $
\begin{align}
s &= (p_c+p_d)^2 , \qquad
t_P=(p_a-p_c)^2\,, \qquad u_P=(p_a-p_d)^2\,,\nn\\
3s_{0P}&\equiv s+t_P+u_P=
 M_{B_s^0}^2+M_{Y}^2+2m_P^2  \,,
\end{align}
where $Y$ denotes the final vector meson $\psi(2S)$ or $X(3872)$, and $P$ represents the pseudoscalar $\pi$ or $K$.
The variables $t_P$ and $u_P$ can
be expressed in terms of $s$ and the helicity angle $\theta$
as
\begin{align}
t_P &= \frac{1}{2} \left[3s_{0P}-s+\kappa_P(s)\cos\theta \right]\,,&
u_P &= \frac{1}{2} \left[3s_{0P}-s-\kappa_P(s)\cos\theta \right]\,, \nn\\
\kappa_P(s) &\equiv \sigma_P
\lambda^{1/2}\big(M_{B_s^0}^2,M_{Y}^2,s\big) \,, & \sigma_P &\equiv
\sqrt{1-\frac{4m_P^2}{s}} \,, \label{eq:tu}
\end{align}
where $\theta$ is defined as the angle between $B_s^0$ and the positive
pseudoscalar meson in the rest frame of the $PP$ system, and
$\lambda(a,b,c)=a^2+b^2+c^2-2(ab+ac+bc)$ is the K\"all\'en triangle function.
We define $\vec{q}$ as the
3-momentum of final vector meson $Y$ in the rest frame of the $B_s^0$ with
\be \label{eq:q} |\vec{q}|=\frac{1}{2M_{B_s^0}}
\lambda^{1/2}\big(M_{B_s^0}^2,M_{Y}^2,s\big) \,. \ee

Using the Lagrangians in
Eqs.~\eqref{LagrangianBspsi2Spipi} and~\eqref{LagrangianBsXpipi}, we can calculate the chiral contact terms for $B_s^0 \to \psi(2S)\pi^+\pi^- (K^+ K^-)$ and $B_s^0 \to X(3872) \pi^+\pi^- (K^+ K^-)$ processes
\begin{align}
M^{\chi, \psi^\prime \pi\pi}(s,\cos\theta)&=0\,, \notag\\
M^{\chi,\psi^\prime KK}(s,\cos\theta)&=-\frac{6}{F^2}\vec{p}_b\cdot\boldsymbol{\epsilon}_{\psi^\prime}\bigg[g_1  p_c\cdot
p_d  +h_1 p_c^0 p_d^0 \bigg] = \sum_{l=0}^\infty M_l^{\chi,\psi^\prime KK}(s) P_l(\cos\theta) \,,  \notag\\
M^{\chi,X \pi\pi}(s,\cos\theta)&=0\,, \notag\\
M^{\chi,X KK}(s,\cos\theta)&=-\frac{6}{F^2}\vec{p}_b\cdot\boldsymbol{\epsilon}_X\bigg[\Big(\tilde{g}_1-\sqrt{2}\tilde{g}_8\Big) p_c\cdot
p_d  +\Big(\tilde{h}_1-\sqrt{2}\tilde{h}_8\Big)p_c^0 p_d^0 \bigg] \notag\\
&= \sum_{l=0}^\infty M_l^{\chi,X KK}(s) P_l(\cos\theta) \,.\label{eq.ContactPi+KRaw}
\end{align}
The amplitude has been partial-wave decomposed, and
$P_l(\cos\theta)$ are the standard Legendre polynomials.
In $B_s^0 \to \psi(2S) \pi^+\pi^-$ and $B_s^0 \to X(3872) \pi^+\pi^-$ processes, isospin conservation combined with Bose
symmetry require the pion pair to be in even angular momentum partial waves. While since $K^+$ and $K^-$ do not belong to the same isospin multiplet, they do not satisfy the Bose symmetry constraint. Therefore, the $P$-wave $K^+K^-$, in fact the $\phi(1020)$, is non-negligible in $B_s^0 \to \psi(2S) K^+ K^-$ and $B_s^0 \to X(3872)K^+ K^-$ processes. However, notice that the $\pi\pi$ system is strongly coupled to the $K\bar{K}$ system via the scalar isoscalar $f_0(980)$ and $f_0(1500)$ resonances, which accounts for the generation of the final states $ \psi(2S) \pi^+\pi^-(X(3872) \pi^+\pi^-)$ from $ \psi(2S) K^+ K^-(X(3872) K^+ K^-)$.
The main focus of this study lies on the treatment of the scalar isoscalar $\pi\pi$-$K\bar{K}$ rescattering, and we will consider the contribution of the $\phi(1020)$ meson to the $B_s^0 \to \psi(2S) K^+ K^-$ and $B_s^0 \to X(3872)K^+ K^-$ processes using Breit--Wigner function. For the even angular momentum partial waves, we only take into account the $S$-
and $D$-wave components in Eq.~\eqref{eq.ContactPi+KRaw}, neglecting the effects of
other partial waves. Explicitly, the projections of $S$- and $D$-waves
of the chiral contact amplitudes read
\begin{align}
M_0^{\chi,\psi^\prime KK}(s)&=-\frac{3}{F^2}\vec{p}_b\cdot\boldsymbol{\epsilon}_{\psi^\prime}
\bigg\{g_1 \left(s-2m_K^2 \right)
+\frac{1}{2} h_1 \bigg[s+\vec{q}^2\Big(1
-\frac{\sigma_K^2}{3} \Big)\bigg]\bigg\}\,, \notag\\
M_2^{\chi,\psi^\prime KK}(s)&=\frac{1}{F^2}\vec{p}_b\cdot\boldsymbol{\epsilon}_{\psi^\prime}
 h_1 \vec{q}^2 \sigma_K^2 \,, \notag\\
M_0^{\chi,X KK}(s)&=-\frac{3}{F^2}\vec{p}_b\cdot\boldsymbol{\epsilon}_X
\bigg\{\Big(\tilde{g}_1-\sqrt{2}\tilde{g}_8\Big) \left(s-2m_K^2 \right)
+\frac{1}{2}\Big(\tilde{h}_1-\sqrt{2}\tilde{h}_8\Big) \bigg[s+\vec{q}^2\Big(1
-\frac{\sigma_K^2}{3} \Big)\bigg]\bigg\}\,, \notag\\
M_2^{\chi,X KK}(s)&=\frac{1}{F^2}\vec{p}_b\cdot\boldsymbol{\epsilon}_X
 \Big(\tilde{h}_1-\sqrt{2}\tilde{h}_8\Big) \vec{q}^2 \sigma_K^2 \,. \label{eq.M0+2Kchiral}
\end{align}

Now we consider the contribution of the $\phi(1020)$ meson to the $B_s^0 \to \psi(2S) K^+ K^-$ and $B_s^0 \to X(3872)K^+ K^-$ decays. In this case, the $K^+K^-$ pair is produced in $P$-wave. The decay amplitudes can be written as~\cite{Wang:2024gsh},
\begin{align}
    M^{\phi,\psi^\prime KK} = g_{B\psi^\prime\phi}g_{\phi KK}\varepsilon^{\mu\nu\rho\sigma}
     \times \epsilon_{\mu}^*(\boldsymbol{p}_{\psi^\prime})p_{{\psi^\prime}\nu}p_{\phi\sigma} \frac{{\rm i}(p_{K^+} - p_{K^-})_{\rho}}{p_\phi^2 - m_{\phi}^2 + {\rm i}m_{\phi}\Gamma_{\phi}} ,\\
     M^{\phi,XKK} = g_{BX\phi}g_{\phi KK}\varepsilon^{\mu\nu\rho\sigma}
     \times \epsilon_{\mu}^*(\boldsymbol{p}_X)p_{X\nu}p_{\phi\sigma} \frac{{\rm i}(p_{K^+} - p_{K^-})_{\rho}}{p_\phi^2 - m_{\phi}^2 + {\rm i}m_{\phi}\Gamma_{\phi}} ,
\label{eq:phiKK}
\end{align}
where $\epsilon_{\mu}^*(\boldsymbol{p}_{\psi^\prime})$ and $\epsilon_{\mu}^*(\boldsymbol{p}_X)$ are the polarization vectors of $\psi(2S)$ and $X(3872)$, respectively. $g_{B\psi^\prime\phi}$, $g_{BX\phi}$, and $g_{\phi KK}$ are the coupling parameters of the vertexes of $B_s^0 \psi(2S)\phi$, $B_s^0X(3872)\phi$, and $\phi KK$, respectively. Using the branching fractions of $\mathcal{B}[{B_s^0 \to \psi(2S)\phi}] = (5.2\pm0.4)\times10^{-4}$, $\mathcal{B}[{B_s^0 \to X(3872)\phi}] = (9.7\pm3.3)\times10^{-5}$, and $\mathcal{B}[{\phi \to K^+K^-}] = (49.1\pm0.5)\%$ in PDG~\cite{ParticleDataGroup:2024cfk}, one can obtain that $g_{B\psi^\prime\phi}^2 = (2.0\pm0.2)\times 10^{-21}\text{MeV}^{-2}$, $g_{BX\phi}^2 = (6.5\pm2.2)\times 10^{-22}\text{MeV}^{-2}$, and $g_{\phi KK}^2 = (20.3 \pm 0.2)$.

\subsection{ Treatment of final-state interactions }

We will use the method developed in Ref.~\cite{Ropertz:2018stk} to consider the strong FSIs between two pseudoscalar mesons. Since the invariant
mass of the pseudoscalar mesons pair reaches about above 1.5 GeV, three-channel ($\pi\pi$ (channel 1), $K\bar K$ (channel 2), and effective $4\pi$ (channel 3) modeled by either $\rho\rho$ or $\sigma\sigma$) FSI will be taken into account for the dominant $S$-wave component. In the low energy region from $\pi\pi$ threshold up to about 1 GeV, the formalism in the method~\cite{Ropertz:2018stk} matches smoothly to that constructed rigorously from dispersion theory. At higher energies, it includes the effect of higher resonances in a way consistent with analyticity.

As elaborated in Ref.~\cite{Ropertz:2018stk}, the strange scalar isoscalar form factor $\Gamma^{s}_i$ can be written as
\begin{align}
	\Gamma^{s}_i=\Omega_{im}\left[1-V_R\Sigma\right]^{-1}_{mn} M_{n}\,,\label{eq::Formalism::Formfactor}
\end{align}
where the channel index $i=1$ $(\pi\pi),2$ $(K\bar{K}),3$ $(4\pi)$.
The Omn\`es function
\begin{align}
	\Omega=\begin{pmatrix}
	\Omega_{11} & \Omega_{12} & 0\\
	\Omega_{21} & \Omega_{22} & 0\\
	0 & 0 & 1
	\end{pmatrix}
\end{align}
is the solution of the homogeneous coupled-channel
unitarity relation
\begin{equation}\label{eq.unitarity2channelhomo}
\textrm{Im}\, \Omega(s)=T_0^{0\ast}(s)\sigma(s) \Omega(s),
\hspace{1cm}  \Omega(0)=\mathbbm{1} \,.
\end{equation}
The three-dimensional matrices $T_0^0(s)$ and $\Sigma(s)$ are represented as
\begin{equation}\label{eq.T00}
T_0^0(s)=
 \left( {\begin{array}{*{3}c}
   \frac{\eta_0^0(s)e^{2i\delta_0^0(s)}-1}{2i\sigma_\pi(s)} & |g_0^0(s)|e^{i\psi_0^0(s)} & 0  \\
  |g_0^0(s)|e^{i\psi_0^0(s)} & \frac{\eta_0^0(s)e^{2i\left(\psi_0^0(s)-\delta_0^0(s)\right)}-1}{2i\sigma_K(s)} & 0\\
  0 & 0 & 0\\
\end{array}} \right),
\end{equation}
and $\sigma(s)\equiv \text{diag}
\big(\sigma_\pi(s)\theta(s-4m_\pi^2),\sigma_K(s)\theta(s-4m_K^2),0 \big)$.
There are three input functions in the $T_0^0(s)$ matrix: the
scalar isoscalar $\pi\pi$ phase shift $\delta_0^0(s)$, and the modulus and phase of the $\pi\pi \to
K\bar{K}$ $S$-wave scattering amplitude $g_0^0(s)=|g_0^0(s)|e^{i\psi_0^0(s)}$.
Note that the numerical result of the $T_0^0(s)$ matrix has been given in Refs.~\cite{Dai:2014lza,Dai:2016ytz} up to $\sqrt{s_0}=1.42$ GeV. We will use the result of the $T_0^0(s)$ matrix in Refs.~\cite{Dai:2014lza,Dai:2016ytz} to extract the information of the phase shift $\delta_0^0(s)$ and the scattering amplitude $g_0^0(s)$. Above $s_0$, the phases
$\delta_0^0(s)$ and $\psi_0^0$ are guided smoothly to
2$\pi$~\cite{Moussallam2000}
\begin{equation}
\delta(s)=2\pi+(\delta(s_0)-2\pi)\frac{2}{1+(\frac{s}{s_0})^{3/2}}\,.
\end{equation}
The inelasticity parameter $\eta_0^0(s)$ in Eq.~\eqref{eq.T00} is related to the modulus $|g_0^0(s)|$
\begin{equation}
\eta_0^0(s)=\sqrt{1-4\sigma_\pi(s)\sigma_K(s)|g_0^0(s)|^2\theta(s-4m_K^2)}\,.
\end{equation}
As the matrix $T_0^0(s)$ captures the information of coupled-channel $\pi\pi$-$K\bar K$ scattering at energies below 1 GeV including the effects of $f_0(500)$ and $f_0(980)$, the potential $V_R$ in Eq.~\eqref{eq::Formalism::Formfactor} predominantly describes the resonances above 1 GeV. The potential $V_R$ is parametrized as
\begin{align}
	\left(V_R\right)_{ij}=  g_{i}^r \,\frac{s}{m_r^2\left(m_r^2-s\right)}\, g_{j}^r\,.
\label{eq::Formalism::subtractedPotential}
\end{align}
Here we only consider one resonance higher than 1 GeV, $f_0(1500)$. The bare resonance mass, $m_r$, and the bare resonance--channel coupling constant, $g_i^r$,
are free parameters that will be determined by fitting to data.
In Eq.~\eqref{eq::Formalism::Formfactor}, the self-energy matrix $\Sigma \equiv G\Omega$ can be calculated as a dispersion integral
\beq
\Sigma_{ij}(s) = \frac{s}{2i \pi}\int \frac{\diff z}{z}\frac{\disc \Sigma_{ij}(z)}{z-s-i\epsilon} \,,
\label{sigmadef}
\eeq
where the discontinuity
\begin{align}
	\disc \Sigma_{ij}(s)=\Omega_{im}^\dagger(s)\, \disc G_{mm}(s) \,\Omega_{mj}(s)\,.
	\label{discSigma}
\end{align}
$G$ is loop function matrix diagonal in the channel space, and it provides the free propagation of the particles in different channels.
For channel $m=1,\,2$, the discontinuity of the loop function matrix element reads
\beq
\disc G_{mm}=2i\sigma_m \,,
\label{G1122def}
\eeq
where $\sigma_m(s) = \sqrt{1-4M_m^2/s}$,
and $M_m$ represents the masses of pion and kaon for channels 1 and 2, respectively.
For the third channel, the finite width of the two broad intermediate ($\rho$ and $\sigma$) mesons need to be considered
\begin{align}
\disc G_{33}^k &=2i \int_{4M_\pi^2}^\infty \diff m_1^2\,\diff m_2^2\,\rho_k(m_1^2)\,\rho_k(m_2^2) \nl
& \qquad\qquad \times \frac{\lambda^{1/2}(s,m_1^2,m_2^2)}{s} \,.
\label{G33def}
\end{align}
Here the spectral density is given as
\begin{align}
	\rho_k(q^2)=\frac{1}{\pi}\,\frac{m_k \Gamma_k(q^2)}{(q^2-m_k^2)^2+m_k^2 \,\Gamma_k^2(q^2)} \,,
\end{align}
with the energy-dependent width
\begin{align}
\Gamma_k(s) &=\frac{\Gamma_k \,m_k}{\sqrt{s}}\,\left(\frac{p_\pi(s)}{p_\pi(m_k^2)}\right)^{2L_k+1}\left(F_R^{(L_k)}(s)\right)^2\,, \nl
p_\pi(s)  &= \frac{\sqrt{s}}{2}\sigma_\pi(s) \,, \label{eq::Decay::width}
\end{align}
where $L_k$ denotes the angular momentum
of the decay with $L_k=1$ and $0$ for the $\rho$ and the $\sigma$, respectively.
The $F_R^{(L)}(s)$ represent barrier factors, and
the parametrization in Refs.~\cite{Blatt:1952ije,LHCb:2012ae} will be used:
\begin{align}
	F_R^{(0)}=1\,,~ F_R^{(1)}=\sqrt{\frac{1+z_0}{1+z}}\,,~ F_R^{(2)}=\sqrt{\frac{9+3z_0+z_0^2}{9+3z+z^2}} \,,\label{eq::Formalism::BlattWeisskopf}
\end{align}
with $z=r_R^2 \,p_\pi^2(s)$, $z_0 = r_R^2 \,p_\pi^2(m_k^2)$, and
the hadronic scale $r_R=1.5\,\GeV^{-1}$. In Eq.~\eqref{eq::Formalism::Formfactor}, $M_i$ describes the transition from
the source to the channel $i$, and the most general ansatz for it reads
\begin{align}
	M_i=c_i+\gamma_i\,s+\cdots -  g_i^r\,\frac{s}{m_r^2-s}\,\alpha_r\,.\label{eq::Formalism::source}
\end{align}
Following Ref.~\cite{Ropertz:2018stk}, the parameters $c_i =\Gamma^{s}_i(0)$ representing the normalizations of the
different form factors can be set as $c_1 =c_3= 0$ and $c_2=1$, and the slope parameters $\gamma_i$ are fixed at 0.
The parameter $\alpha_r$ describing the resonance--source coupling is an additional free parameter.

For $B_s^0 \to \psi(2S) \pi^+\pi^-(X(3872) \pi^+\pi^-)$ processes, the dominant generation mechanism is $B_s^0 \to \psi(2S) K\bar{K}(X(3872) K\bar{K})$ with
kaons rescattering to a pion pair via the scalar isoscalar $f_0(980)$ and $f_0(1500)$ resonances, therefore we only consider the $S$-wave contribution. In the fitting to the experimental data, we use the following form for the $\pi\pi$ invariant mass distribution of $B_s^0 \to \psi(2S) \pi^+\pi^-(X(3872) \pi^+\pi^-)$
\begin{equation}
\frac{\diff\Gamma}{\diff m_{\pi\pi}} =
C\frac{\sqrt{s}\sigma_\pi \vec{q} |2/\sqrt{3} M_0^{\chi,\psi^\prime KK(XKK)} \Gamma^{s}_1 |^2 }{192\pi^3 M_{B_s}^2 }\,,\label{eq.pipimassdistribution}
\end{equation}
where $C$ is the normalization factor since the experimental data are given in events. With the branching ratio of $ \mathcal{B}[B_s^0 \to \psi(2S)\pi^+\pi^-] = (6.9 \pm 1.2) \times 10^{-5}$~\cite{ParticleDataGroup:2024cfk}, we can extract the value of the normalization factor $C = (8.28 \pm 1.44)\times 10^{17}$. Note that the normalization factor $C$ is the same for the
two processes $B_s^0 \to \psi(2S) \pi^+\pi^-$ and $B_s^0 \to X(3872) \pi^+\pi^-$.

For $B_s^0 \to \psi(2S) K^+ K^-(X(3872) K^+ K^-)$ processes, we consider both the $S$- and $D$-wave contributions and the $\phi(1020)$-exchange contribution, and the $K^+ K^-$ invariant mass distribution reads
\begin{equation}
\frac{\diff\Gamma}{\diff m_{KK}} =C\int_{-1}^1
\frac{\sqrt{s}\sigma_K \vec{q} \big( | M_0^{\chi,\psi^\prime KK(XKK)} \Gamma^{s}_2 +M_2^{\chi,\psi^\prime KK(XKK)}P_2(\cos\theta)  |^2 +|M^{\phi,\psi^\prime KK(XKK)}  |^2 \big) }{192\pi^3 M_{B_s}^2 } \diff
\cos\theta \,.\label{eq.KKmassdistribution}
\end{equation}

\section{Phenomenological discussion}\label{pheno}

In this work we perform fits taking into account the experimental data sets of the $\pi\pi$ invariant mass distributions of $B_s^0 \to \psi(2S) \pi^+\pi^-$ and $B_s^0 \to X(3872) \pi^+\pi^-$ processes~\cite{LHCb:2023reb}, the $K\bar{K}$ invariant mass distributions of $B_s^0 \to X(3872) K^+ K^-$ process~\cite{LHCb:2020coc}, and the ratio of branch fractions
$\mathcal{B}[B_s^0 \to X(3872)(K^+K^-)_{{\rm non-}\phi}]/ \mathcal{B}[B_s^0 \to X(3872)\pi^+ \pi^-]$~\cite{ParticleDataGroup:2024cfk}.
There are nine free parameters in our fits: $g_1$, $h_1$, the combinations of the coupling constants $\tilde{g}_1-\sqrt{2}\tilde{g}_8$ and $\tilde{h}_1-\sqrt{2}\tilde{h}_8$, $m_{r}$, $g_{1,2,3}^r$, and $\alpha_{r}$. The parameters $g_1$ and $h_1$ correspond to the coupling constants in the $B_s^0 \to \psi(2S)\pi^+\pi^- (K^+ K^-)$ amplitudes, and the combinations of the coupling constants $\tilde{g}_1-\sqrt{2}\tilde{g}_8$ and $\tilde{h}_1-\sqrt{2}\tilde{h}_8$ are corresponding parameters in the $B_s^0 \to X(3872) \pi^+\pi^- (K^+ K^-)$ amplitudes given in Eq.~\eqref{eq.M0+2Kchiral}.
$m_{r}$, $g_{1,2,3}^r$, and $\alpha_{r}$ are related to the bare resonance ($f_0(1500)$) mass, the bare resonance--channel coupling constant, and the resonance--source coupling, respectively. We perform two fits, in which Fit~I assumes the third channel is dominated by $\rho\rho$, and Fit~II assumes the third channel is dominated by $\sigma\sigma$.

\begin{table}
\caption{\label{tablepar} The parameter results from the fits of the $\pi\pi$ invariant mass distributions of $B_s^0 \to \psi(2S) \pi^+\pi^-$ and $B_s^0 \to X(3872) \pi^+\pi^-$ processes, the $K\bar{K}$ invariant mass distributions of $B_s^0 \to X(3872) K^+ K^-$ process, and the ratio of branch fractions
$\mathcal{B}[B_s^0 \to X(3872)(K^+K^-)_{{\rm non-}\phi}]/ \mathcal{B}[B_s^0 \to X(3872)\pi^+ \pi^-]$. Fits~I and II assume the third channel is dominated by $\rho\rho$ and $\sigma\sigma$, respectively. }
\renewcommand{\arraystretch}{1.2}
\begin{center}
\begin{tabular}{l|cc}
\toprule
         & Fit~I
         & Fit~II\\
\hline
$g_1 \times 10^{10} ~[\text{GeV}^{-1}]$   &    $ 2.77\pm 0.01$  &   $ 2.91\pm 0.01$   \\
$h_1 \times 10^{10} ~[\text{GeV}^{-1}]$   &    $ 2.33\pm 0.01$  &   $ 2.43\pm 0.01$   \\
$(\tilde{g}_1-\sqrt{2}\tilde{g}_8)  \times 10^{10} ~[\text{GeV}^{-1}]$   &    $ 1.88\pm 0.01$  &   $ 1.95\pm 0.01$   \\
$(\tilde{h}_1-\sqrt{2}\tilde{h}_8) \times 10^{10} ~[\text{GeV}^{-1}]$   &    $ 0.91\pm 0.01$  &   $ 0.96\pm 0.01$   \\
$m_r  ~[\text{GeV}]$   &    $ 1.45\pm 0.10$  &   $ 1.47\pm 0.10$   \\
$g_1^r  ~[\text{GeV}]$   &    $ -0.14\pm 0.01$  &   $ -0.08\pm 0.01$   \\
$g_2^r  ~[\text{GeV}]$   &    $ -0.34\pm 0.02$  &   $ -0.18\pm 0.03$   \\
$g_3^r  ~[\text{GeV}]$   &    $ 0.91\pm 0.05$  &   $ 0.89\pm 0.04$   \\
$\alpha_r  ~[\text{GeV}^{-1}]$   &    $ 1.50\pm 0.05$  &   $ 2.49\pm 0.03$   \\
\hline
 ${\chi^2}/{\rm d.o.f.}$ &  $\frac{150.2}{(117-9)}=1.39$  &  $\frac{130.6}{(117-9)}=1.21$     \\
\botrule
\end{tabular}
\end{center}
\renewcommand{\arraystretch}{1.0}
\end{table}

\begin{figure}
\centering
\includegraphics[width=\linewidth]{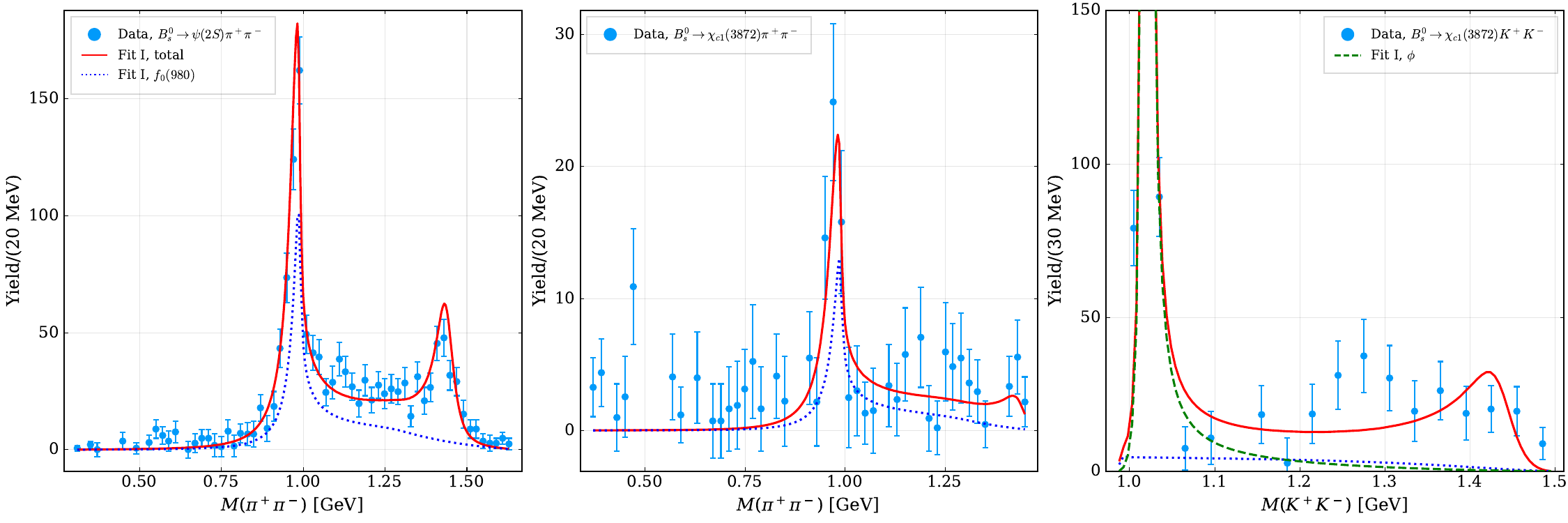}
\includegraphics[width=\linewidth]{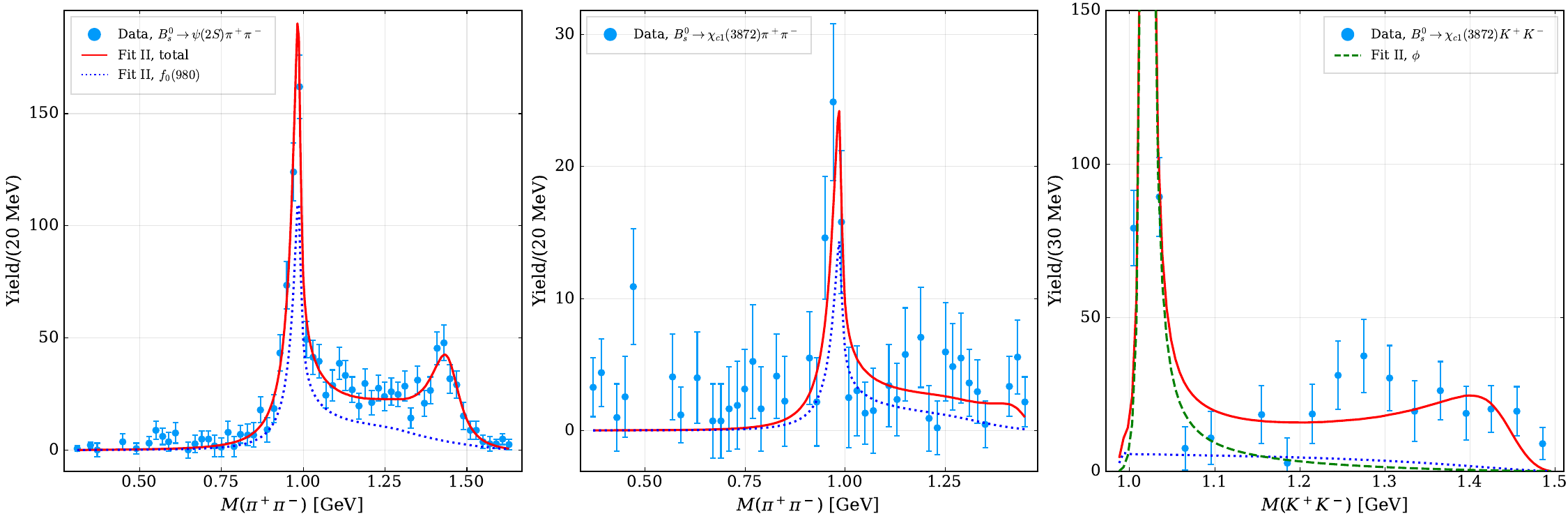}
\caption{Fit results of the $\pi\pi$ invariant mass distributions of $B_s^0 \to \psi(2S) \pi^+\pi^-$ (left) and $B_s^0 \to X(3872) \pi^+\pi^-$ processes (middle), and the $K\bar{K}$ invariant mass distributions of $B_s^0 \to X(3872) K^+ K^-$ process (right) for Fits~I (top) and II (bottom). The red solid lines represent the best fit results, while the blue dotted and green dashed lines correspond to the contributions from the $f_0(980)$ only and the $\phi(1020)$ meson, respectively. 
}\label{fig.FitResults}
\end{figure}

The fit results of invariant mass spectra are shown as red solid lines in Fig.~\ref{fig.FitResults}. The fit result of the branching fractions ratio
of $\mathcal{B}[B_s^0 \to X(3872)(K^+K^-)_{{\rm non-}\phi}] /\mathcal{B}[B_s^0 \to X(3872)\pi^+ \pi^-)$ is $1.38 \pm 0.23$ and $1.37 \pm 0.21$ in Fit~I and II, respectively, which agrees with the value of $2.05 \pm 1.16$ from PDG~\cite{ParticleDataGroup:2024cfk}.
The fitted parameters as well as the $\chi^2/\text{d.o.f.}$ are given in Table~\ref{tablepar}.

If we take the same coupling constants $g_1$ and $h_1$ in Table~\ref{tablepar} for the $B_s^0 \to J/\psi \pi^+\pi^-$ decay, we obtain $\Gamma[B_s^0 \to J/\psi f_0(980) \to J/\psi \pi^+\pi^-] = (6.5 \pm 0.8) \times 10^{-14}$ MeV and $(7.0 \pm 1.0) \times 10^{-14}$ MeV for Fits~I and II, respectively, which are in agreement with the value of $(5.4 \pm 0.7)\times 10^{-14}$ MeV quoted in the PDG~\cite{ParticleDataGroup:2024cfk}. This indicates the universality of the coupling constants for producing charmonium states in the $B^0_s$ decays. While one observes that the values of the combinations of the coupling constants $(\tilde{g}_1-\sqrt{2}\tilde{g}_8)$ and $(\tilde{h}_1-\sqrt{2}\tilde{h}_8)$ in $B_s^0 \to X(3872) \pi^+\pi^- (K^+ K^-)$ are about half of magnitude smaller than  $g_1$ and $h_1$ in $B_s^0 \to \psi(2S)\pi^+\pi^- (K^+ K^-)$, respectively. This indicates that production mechanisms of $\psi(2S)$ and $X(3872)$ in the $B^0_s$ decays are different. Indeed, the $\bar{D}D^\ast$ rescattering mechanism~\cite{Artoisenet:2010va,Braaten:2019yua} would give a large contribution to $X(3872)$ production in the $B_s^0$ decays.

For the $\pi\pi$ mass spectra in $B_s^0 \to \psi(2S) \pi^+\pi^-$ process, one observes that the peaks around 1 GeV and 1.45 GeV due to the presences of the $f_0(980)$ and $f_0(1500)$, respectively, are described well in both Fits~I and II. For the $\pi\pi$ mass spectra in $B_s^0 \to X(3872) \pi^+\pi^-$ process, the peak around 1 GeV is reproduced well, while the peak due to the $f_0(1500)$ is suppressed by the small phase space. In Fig.~\ref{fig.FitResults}, we also plot the contributions from the $f_0(980)$ only by blue dotted lines. One observes that for the peak around 1 GeV in the $\pi\pi$ mass spectra in both $B_s^0 \to \psi(2S) \pi^+\pi^-$ and $B_s^0 \to X(3872) \pi^+\pi^-$ processes, the blue dotted lines only contribute about a half of height. The observation can be confirmed by comparing the branching fractions. Using the fitted coupling constants in Table~\ref{tablepar}, we obtain
\bea
    \frac{\mathcal{B}[B_s^0 \to \psi(2S)(f_0(980)\to \pi^+ \pi^-)]}{\mathcal{B}[B_s^0 \to \psi(2S)\pi^+ \pi^-)]}=& 0.41 \pm 0.05 ~ \text{(Fit I)}/ 0.45 \pm 0.06 ~\text{(Fit II)}\,,\nn\\
    \frac{\mathcal{B}[B_s^0 \to X(3872)(f_0(980)\to \pi^+ \pi^-)]}{\mathcal{B}[B_s^0 \to X(3872)\pi^+ \pi^-)]}=& 0.46 \pm 0.06 ~ \text{(Fit I)}/ 0.51 \pm 0.0.06 ~\text{(Fit II)}\,,
\label{eq:ratiooff0980overtotalpipi}
\eea
which shows that the $\pi^+\pi^-$ contribution from $f_0(980)$ only accounts for approximately half of the total. Therefore we conclude that in $B_s^0 \to \psi(2S)[X(3872)] \pi^+\pi^-$ decays, the interference between the contribution of $f_0(980)$ and the contribution of $f_0(1500)$ is nonnegligible.

Note that the $B_s^0$ is very close to the $X(3872) f_0(1500)$ threshold, and therefore the phase space of $B_s^0 \to X(3872) f_0(1500)$ is much smaller than that of $B_s^0 \to X(3872) f_0(980)$. While as shown in
Fig.~\ref{fig.FitResults}, the contribution of $B_s^0 \to X(3872) f_0(1500) \to X(3872) K^+K^-$ dominate in $B_s^0 \to X(3872)(K^+K^-)_{{\rm non-}\phi}$.
Indeed, using the fitted coupling constants in Table~\ref{tablepar}, we obtain
\bea
    \frac{\mathcal{B}[B_s^0 \to X(3872)(f_0(980)\to K^+ K^-)]}{\mathcal{B}[B_s^0 \to X(3872)(K^+ K^-)_{{\rm non-}\phi}]}=& 0.22 \pm 0.03 ~ \text{(Fit I)}/ 0.24 \pm 0.04 ~\text{(Fit II)}\,,
\label{eq:ratiooff0980overtotalpipi}
\eea
which means that the $S$-wave $K^+ K^-$ contribution from $f_0(980)$ is small compared with the contribution from $f_0(1500)$.
Since the mechanism $B_s^0  \to X(3872) K^+ K^-$ with the $S$-wave kaons rescattering to a pion pair is an important contribution to $B_s^0 \to  X(3872)\pi^+\pi^-$, this is consistent with the observation we made above, namely that effect of $f_0(1500)$ is nonnegligible in the $B_s^0 \to X(3872)\pi^+ \pi^-$ decay.
Notice that Ref.~\cite{Wang:2024gsh} also phenomenologically investigated $B_s^0 \to X(3872)[\psi(2S)]\pi^+ \pi^-(K^+ K^-)$ decays.
Within the chiral unitary approach, the $f_0(980)$ yielded from dynamical generation was considered for both $B_s^0 \to X(3872)\pi^+ \pi^-(K^+ K^-)$ and $B_s^0 \to \psi(2S)\pi^+ \pi^-(K^+ K^-)$ processes. While the effect of $f_0(1500)$ was only considered for the $B_s^0 \to \psi(2S)\pi^+ \pi^-(K^+ K^-)$ processes in Ref.~\cite{Wang:2024gsh}. In our scheme, the effects of $f_0(980)$ and $f_0(1500)$ are considered for both $B_s^0 \to X(3872)\pi^+ \pi^-(K^+ K^-)$ and $B_s^0 \to \psi(2S)\pi^+ \pi^-(K^+ K^-)$ processes in a way fulfilling unitary and analyticity. Another merit of our study is that, instead of only fitting the $\pi^+\pi^-$ invariant mass spectra of $B_s^0 \to \psi(2S) \pi^+\pi^-$ in Ref.~\cite{Wang:2024gsh}, we simultaneously take into account the experimental data of $B_s^0 \to \psi(2S) \pi^+\pi^-$ and $B_s^0 \to X(3872) \pi^+\pi^- (K^+ K^-)$ processes in our fitting.

\begin{figure}
\centering
\includegraphics[width=\linewidth]{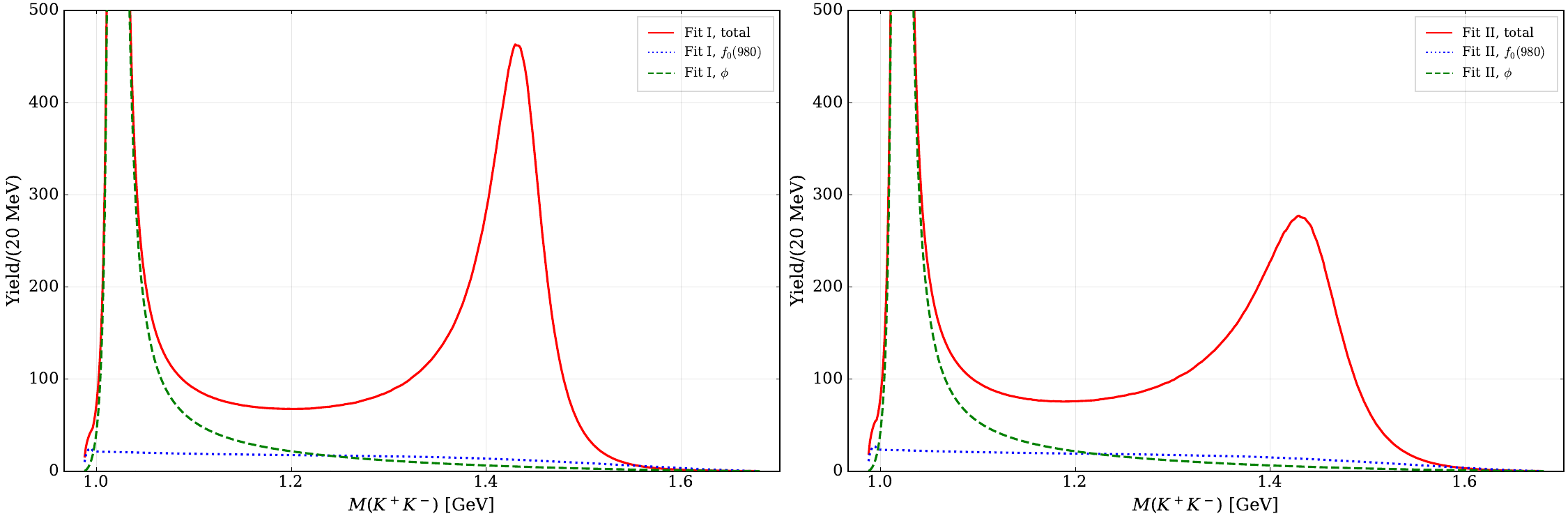}
\caption{Theoretical prediction of the $K\bar{K}$ invariant mass distributions of $B_s^0 \to \psi(2S)(K^+K^-)_{{\rm non-}\phi}$ process for Fits~I (left) and II (right). The red solid lines represent the best fit results, while the blue dotted and green dashed lines correspond to the contributions from the $f_0(980)$ only and the $\phi(1020)$ meson, respectively. 
}\label{fig.KKspectraofpsi2SKK}
\end{figure}

Using the fit parameters given in Table~\ref{tablepar}, we can
predict the ratio of branch fractions
$\mathcal{B}[B_s^0 \to \psi(2S)(K^+K^-)_{{\rm non-}\phi}]/ \mathcal{B}[B_s^0 \to \psi(2S)\pi^+ \pi^-]$, as well as the $K\bar K$ invariant mass
distribution of $B_s^0 \to \psi(2S)K^+K^-$. The prediction of the ratio of branch fractions is
\bea
    \frac{\mathcal{B}[B_s^0 \to \psi(2S)(K^+K^-)_{{\rm non-}\phi}]}{\mathcal{B}[B_s^0 \to \psi(2S)\pi^+ \pi^-)]}=& 2.22 \pm 0.30 ~ \text{(Fit I)}/ 2.09 \pm 0.29 ~\text{(Fit II)}\,,
\label{eq:ratioofpsi2SKKoverpsi2Spipi}
\eea
and the dikaon invariant mass
spectrum is given in Fig.~\ref{fig.KKspectraofpsi2SKK}.
The contribution of the $f_0(980)$ shows a rapid rise in the
near-threshold region of the $K\bar K$ invariant mass distribution, in line with
the peak around $1\GeV$ in the $\pi\pi$ invariant mass distributions in Fig.~\ref{fig.FitResults}. Also one observes that the contribution of $B_s^0 \to \psi(2S) f_0(1500) \to \psi(2S) K^+K^-$ is much larger than
the contribution of $B_s^0 \to \psi(2S) f_0(980) \to \psi(2S) K^+K^-$. These predictions encourage future
experimental measurements in this channel.

\section{Conclusions}
\label{conclu}

We have studied the processes $B_s^0 \to \psi(2S) \pi^+\pi^-$ and $B_s^0 \to X(3872) \pi^+\pi^- (K^+ K^-)$. The strong $S$-wave coupled-channel FSI has been considered using the parametrization developed in Ref.~\cite{Ropertz:2018stk}, which matches a rigorous dispersive description at low and agrees with the phenomenological success of a unitary and analytic isobar model beyond. Through simultaneously fitting the data of the $\pi^+\pi^-$ invariant mass spectra of $B_s^0 \to \psi(2S) \pi^+\pi^-$, the $\pi^+\pi^-$ and $K^+ K^-$ invariant mass spectra of $B_s^0 \to X(3872) \pi^+\pi^- (K^+ K^-))$, and the ratio of branch fractions $\mathcal{B}[B_s^0 \to X(3872)(K^+K^-)_{{\rm non-}\phi}]/ \mathcal{B}[B_s^0 \to X(3872)\pi^+ \pi^-)$, we determine the couplings of the $B_s^0 \psi(2S) PP$ vertex and the $B_s^0 X(3872) PP$ vertex, the $f_0(1500)$ mass, the $f_0(1500)$--channel coupling constants, and the $f_0(1500)$--source coupling. If we take the same couplings of the $B_s^0 \psi(2S) PP$ vertex to calculate the $B_s^0 \to J/\psi \pi^+\pi^-$ decay, the theoretical result of $\Gamma[B_s^0 \to J/\psi f_0(980) \to J/\psi \pi^+\pi^-]$ agrees with the value in the PDG, which shows the universality of the coupling constants for producing charmonium states in the $B^0_s$ decays. While the couplings of the $B_s^0 X(3872) PP$ vertex are about half of magnitude smaller than the couplings of the $B_s^0 \psi(2S) PP$ vertex, which indicates that the $X(3872)$ is not a pure charmonium state. Furthermore, we find that the $f_0(1500)$ plays an important role in the $B_s^0 \to \psi(2S) \pi^+\pi^-$ and the $B_s^0 \to X(3872) \pi^+\pi^- (K^+ K^-)$ processes, though the phase space of $B_s^0 \to X(3872) f_0(1500)$ is tiny. We also predict the ratio of branch fractions $\mathcal{B}[B_s^0 \to \psi(2S)(K^+K^-)_{{\rm non-}\phi}]/ \mathcal{B}[B_s^0 \to \psi(2S)\pi^+ \pi^-]$ and the $K\bar K$ invariant mass
distribution of $B_s^0 \to \psi(2S)K^+K^-$, which can be tested by future experimental measurements.

\end{document}